\documentclass[12pt]{iopart}
\usepackage{iopams}  
\usepackage{cite} % macht [2,3,4] zu [2-4]
\usepackage{epsfig}
\usepackage{caption}
\usepackage{psfig,psfrag}
\usepackage{amssymb,bbm}

\newcommand{\ket}[1]{\, | #1 \rangle}

\newcommand{\mb}[1]{{\rm #1}}

\newcommand{\Hnorm}{\mb{H}}

\newcommand{\om}{\omega _0}

\newcommand{\Pt}{P_{\rm ion}(t)}

\begin{document}

\title[Atomic conductance]{Signatures of Anderson localization in the 
ionization rates of periodically driven Rydberg states} 

\author{Sandro Wimberger and Andreas Buchleitner} 

\address{Max-Planck-Institut f\"ur Physik komplexer Systeme,
N\"othnitzer Str. 38, D-01187 Dresden}
\address{e-mail: saw@mpipks-dresden.mpg.de}

%\ead{saw@mpipks-dresden.mpg.de}

\begin{abstract}
We provide a statistical characterization of the ionization 
yield of one-dimensional, periodically driven 
Rydberg states of atomic hydrogen, in the spirit of 
Anderson localization theory. We find excellent agreement
with predictions for the conductance across an Anderson localized,
quasi one-dimensional, disordered wire, in the semiclassical limit of highly
excited atomic initial states. For the moderate atomic excitations typically 
encountered in state of the art laboratory experiments, finite-size 
effects induce significant deviations from the solid-state picture. However,
large scale fluctuations of the atomic conductance prevail and are robust
when averaged over a finite interval of driving field amplitudes, as
inevitably done in the experiment.
\end{abstract}


% Uncomment for Submitted to journal title message\usepackage{amsmath}

%\maketitle

\section{Introduction}
\label{intro}
Coherent quantum transport on a mesoscopic scale is the origin of 
many intriguing transport phenomena in complex systems \cite{houches95}.
The somewhat vague attribute ``complex'' summarizes a multitude of more 
specific physical situations: complex dynamics can be generated by disorder,
by many-particle interactions, and by dynamical chaos, to name a few.

Arguably one of the most prominent and most fundamental coherence effects
in complex quantum transport is Anderson localization \cite{And58}, 
the quantum suppression
of conductance across a disordered, quasi one-dimensional solid-state lattice.
Viewed as a scattering problem, it essentially manifests itself 
in exponentially small
transmission probabilities from input to output of the sample, as a consequence
of multiple scattering events (with finite reflection and transmission 
coefficients) at randomly placed scattering sites along the lattice. This 
naturally generates a multitude 
of transmission
amplitudes which have to be summed up coherently on output. If their 
individual phases have been randomized by
the disordered lattice potential, they will tend
to interfere destructively. 
In terms of electronic eigenfunctions, Anderson localization enforces
their exponential localization on the lattice 
domain. The degree of localization is characterized by the
localization length $\xi$, which should be compared to the sample size $L$,
in order to allow predictions on the conductance across the sample
\cite{KM93}. 
Complexity is brought about in this problem by two 
components: (a) the large number of interfering transmission amplitudes, and 
(b) the disordered lattice potential which breaks the translational invariance
(which otherwise would reduce the complexity introduced by (a) through some
kind of Bragg condition). 

As mentioned before, complexity can have different causes, and therefore 
a random potential is not necessary to enforce vanishing total transition 
amplitudes or exponentially localized eigenfunctions -- any mechanism which 
equidistributes the phases of the individual interfering amplitudes will do.
In particular, dynamical chaos can substitute for disorder in quantum systems
with a well-defined classical counterpart, and more generally, if such an 
analogy is unavailable, deterministic quantum systems with largely broken 
symmetries or non-perturbatively coupled degrees of freedom (displaying 
quantum chaos) can replace 
the simple scenario of a disordered lattice 
(there is a caveat concerning 
the dimensionality of the dynamics, but we shall restrain
here to effectively one-dimensional systems) \cite{CCFI79,FGP82,CCS84,CC87}. 
The only further ingredient 
required for such systems to mimic Anderson's scenario is a sufficiently high
density of states, such that a sufficient number of transition 
amplitudes can interfere. At the quantum-classical interface, this corresponds
to sufficiently small values of the effective Planck constant 
$\hbar_{\rm eff}$, which is determined (via the uncertainty principle)
by the comparison of $\hbar$ to the typical scales of the given problem, 
measured in canonical action-angle variables \cite{CC87}.

Besides simple billiard shaped cavities, which are of some relevance in the 
context of microdisc lasers \cite{SJNS00}, 
strongly perturbed atomic \cite{CC87,DB94} and molecular \cite{LW97}
systems are 
perfect candidates to study signatures of Anderson localization in 
quantum systems without disorder. The simplest
(though realistic) representative of the latter is the hydrogen atom
exposed to electromagnetic fields \cite{CC87}, with the atom initially 
prepared in a 
Rydberg level of principal quantum number $n_0\gg 1$, and a driving 
field frequency
$\omega\sim n_0^{-3}$ near resonant with the atomic transition 
$n_0\rightarrow n_0+1$, 
i.e., within
the microwave range. Such choice of the relevant parameters satisfies both
general requirements stated above: on the one hand, the ionization potential
of a Rydberg state $\ket {n_0}$ (we neglect the angular degree of freedom
in our present treatment) requires the (net) absorption of 
approximately $N\simeq 1/2n_0^2\omega\sim n_0$ 
photons to establish a transition to 
the atomic continuum. If, much as in the solid-state problem, each atomic 
bound state $n>n_0$ (with energy $E=-1/2n^2$) which is quasi resonantly 
coupled to the initial state
(i.e., $1/2n_0^2-1/2n^2=m\omega +\delta$, $m$ integer, and 
$\delta\ll\omega$ the detuning from resonance) plays the role of a single 
scatterer of the lattice, then
a large number of transition amplitudes between 
$\ket {n_0}$ and the continuum become available (notice that emission
events will eventually couple states with $n<n_0$ as well). 
This number rapidly increases with $N$, which 
therefore plays the role of the ``atomic sample size'', 
in analogy to the length $L$ of a solid-state sample \cite{BG98}. 
On the other hand, due to the nonlinearity of the 
Coulomb potential, the detuning $\delta$, which determines the coupling 
strength between quasi resonantly coupled states, will be effectively 
randomized (much alike the simple generation of random numbers by a mod 
operation \cite{PT94}), and this accounts for randomizing the phases of the 
various  
transition amplitudes which mediate the ionization process \cite{CGS90}. 
Consequently, the
general scenery for Anderson localization to occur -- this time on the energy 
axis rather than along the lattice -- is set. In addition, a perfect
classical analogue exists for the driven hydrogen atom: It is
well-established that ionization is brought about by classical chaos,
in the specified
parameter range, since the quasi resonant coupling of sequences of Rydberg 
states described above destroys the good quantum numbers of the problem,
which is synonymous to nonintegrability on the classical level 
\cite{CC87,DB94}. 

Indeed, the above analogy between charge transport through disordered solids
and the ionization of Rydberg states by microwave fields has been identified
approximately 20 years ago \cite{CCFI79,FGP82,CCS84,CC87}. 
Baptized ``dynamical localization'' (to stress
its origin in dynamical chaos rather than in disorder) it has been 
qualitatively 
demonstrated by several independent experimental groups, on various atomic 
species \cite{GSMKR88,BCGS89,BBGSSW91,ABMW91,NGG00}. 
A theoretical framework -- known as ``photonic localization 
theory'' \cite{CC87} -- which 
relies on an ingenious mixture of crude approximations on the
atomic side and deep physical intuition on the statistical side, provides 
explicit expressions for the mean of the localization length 
\begin{equation}
\langle \xi \rangle = 3.33 F_0^2 \omega_0^{-10/3} n_0^2,
\label{locallength}
\end{equation}
and for the sample size 
\begin{equation}
N = \frac{n_0}{2\omega_0} \left( 1 - \frac{n_0^2}{n_c^2}\right),
\label{samplelength}
\end{equation}   
where the factor in parenthesis in (\ref{samplelength}) accounts for a 
shift of the ionization threshold to the finite value $n_c<\infty$, induced by 
experimentally unavoidable stray electric fields \cite{ABMW91}. 
$\omega_0=\omega n_0^3$ and $F_0=Fn_0^4$ are the frequency and amplitude 
scaled with respect to the classical Kepler frequency, and to the Coulomb 
field 
amplitude along the unperturbed classical Rydberg orbit, respectively 
\cite{CC87}.

According to the scaling theory of localization \cite{AALR79}, 
$\xi$ fluctuates around 
$\langle\xi\rangle$ for different realizations of the sample, at finite $N$, 
and tends to the non-fluctuating, sample 
independent value $\langle\xi\rangle$ only for $N\rightarrow\infty$, with 
the statistical distribution of $\xi$ completely determined by the 
localization parameter 
\begin{equation}
\ell \equiv \frac{\langle\xi\rangle}{N}\simeq
\frac{6.66 F_0^2 n_0}{\omega_0^{7/3}}\left(1-\frac{n_0^2}{n_c^2}\right)^{-1}.
\label{lparameter}
\end{equation}
By virtue of this last expression, different realizations of the same value
of $\ell$ can be realized, at fixed $n_0$, by simultaneously tuning 
$F_0$ and $\omega_0$ over a finite interval.
Since, for an exponentially localized wave function on the energy axis, the 
population close to threshold is $\sim\exp(-2N/\xi)$, 
the ``atomic conductance'' $g$, and this is nothing but the total
transition probability to the atomic continuum, should then reflect 
the fluctuations of $\xi^{-1}$ in {\em exponentially enhanced} fluctuations 
via
\begin{equation}
g\sim\exp(-2N/\xi).
\label{atcond}
\end{equation}
Consequently, Anderson localization of the electronic bound-space
population of a periodically driven Rydberg state implies
large scale fluctuations of $g$ and also of the total ionization yield, under
changes of $\omega_0$, at fixed values of $\ell$ and $n_0$. 

This latter prediction has been verified in a recent publication \cite{BG98}
on the conductance of periodically driven one-dimensional Rydberg states of
atomic hydrogen. More precisely, \cite{BG98} demonstrated 
the lognormal distribution of $g$ (which follows
from a normal distribution of $\xi^{-1}$, via (\ref{atcond})), and the 
approximately linear dependence of ${\rm Var}(\ln g)$ on 
$\langle \ln g\rangle $ \cite{PZIS90}, 
for the single value $n_0=70$ of the principal quantum
number. However, the following highly relevant questions remained unaddressed:
\begin{itemize}
\item Sample size $N$ and localization parameter $\ell$ explicitely depend on
$n_0$, and therefore on $\hbar_{\rm eff}\sim n_0^{-1}$ (the latter relation is
a direct consequence of
the scale invariance of the classical equations of motion of the driven 
Rydberg electron \cite{CC87}). 
In the light of our qualitative discussion above, 
$N$ and therefore $n_0$ must not be too small for the Anderson picture to 
prevail in the atomic ionization process, since otherwise not enough 
transition amplitudes with quasi random phases will contribute to the total 
ionization yield. Hence, are we able to detect significant 
deviations from the predictions of Anderson's model in the atomic problem, 
for smaller values of $n_0$?
\item Which are the smallest values of $n_0$ for which the dominant 
signatures of 
Anderson localization remain detectable in the ionization process?
\item Can we confirm the linear dependence of ${\rm Var}(\ln g)$ on 
$\langle \ln g\rangle $ for variable values of $n_0$?
\item Under which conditions are the predicted fluctuations of the atomic
conductance experimentally observable?
\end{itemize}
The present contribution attempts to answer these questions. 

The paper is organized as follows: Section \ref{theory} summarizes our 
theoretical/numerical approach to the atomic problem at hand, and 
introduces our definition of the atomic conductance. In section
\ref{atomfluct} we investigate 
the statistical properties of the atomic conductance, and particularly 
their dependence on the principal quantum number $n_0$, 
which explicitely enters equations~(\ref{locallength}-\ref{lparameter}). 
Section \ref{conclusions} concludes the paper, with a discussion of the
experimental implications of our results.

\section{Theoretical background}
\label{theory}

The Hamiltonian of a hydrogen atom exposed to an electromagnetic field 
polarized along the $z$-axis reads, in atomic units: 
\begin{equation}
 \Hnorm   =  \frac{1}{2}\vec{p}{\,^2} -\frac{1}{r} -  \frac{Fp_z}{\omega}
\sin(\omega t).
\label{hamilton2}
\end{equation}
Here, the dipole approximation in the velocity gauge was used, we 
dropped the ponderomotive energy shift,
assumed an infinite mass of the nucleus, and neglected 
relativistic effects \cite{BDG95}. 
In the following, we shall furthermore restrict 
configuration space to the single dimension defined by the field polarization
axis, which results in the Hamiltonian
\begin{equation}
\fl \Hnorm =\frac{1}{2}{p_z}^{\,2}+V(z)-\frac{Fp_z}{\omega}\sin(\omega t),
~~~\mbox{with}~~~V(z) = \left\{ 
\begin{array}{r@{\quad , \quad}l}
-\frac{1}{z} & z>0 \\ \infty & z \leq 0.
\end{array} \right.
\label{hamilton6}
\end{equation}
This approximation was chosen to keep the numerical effort necessary for 
sampling sufficient statistical data within reasonable bounds, and is also 
justified for atoms initially prepared in extremal parabolic states which are
quasi one-dimensional eigenstates of the unperturbed hydrogen atom 
\cite{KL95,BD97}. 
Even for real three-dimensional atomic initial states with low angular 
momentum quantum numbers has this one-dimensional model been shown to yield 
quantitatively satisfactory results, within a certain parameter regime
\cite{BDG95,BD97}. 
Nonewithstanding, an extension of our present work to the real
three-dimensional world remains clearly desirable and will bear further 
surprises, but is at present an extremely expensive enterprise which 
saturates the largest computer facilities currently available. For the time 
being, despite their restricted range of predictive power, our subsequent 
results exhibit enough novel phenomena which shed new light on the atomic 
ionization process, and indicate the road to follow in future 3D calculations.

In order to extract ionization yields from the time-periodic Hamiltonian
(\ref{hamilton6}), we exploit the Floquet theorem and diagonalize the Floquet 
Hamiltonian ${\mathcal H}= \Hnorm -\rmi \partial_{t}$ 
in a Sturmian basis, after complex dilation \cite{BDG95}. 
This provides direct access to the poles of the 
resolvent of ${\mathcal H}$, and hence an exact representation of the 
Green's function and of the associated time evolution operator of our
problem. The latter finally leads 
(after an average over the initial phase of the driving field) 
to the following expression for the atomic ionization yield 
$P_{\rm ion}(t)$ as a function of the interaction time $t$ \cite{BDG95}:
\begin{equation}
P_{\rm ion}(t)=1-\sum_{\epsilon}w_{\epsilon}\exp(-\Gamma_{\epsilon}t),\ t>0.
\label{ionprob}
\end{equation}
The sum runs over a single Floquet zone of length $\omega$ on the energy axis
\cite{BDG95}, the $\Gamma_{\epsilon}$
represent the ionization rates of individual Floquet eigenstates 
$\ket {\epsilon}$ of the atom in
the field, and the $w_{\epsilon}$ are their weights 
in the decomposition of the atomic initial state $\ket {n_0}$
over the Floquet basis. Note that for $n_0\simeq 40\ldots 100$ approximately
$50\ldots 120$ Floquet states contribute with 
non-vanishing $w_{\epsilon}$ to the sum in (\ref{ionprob}). Therefore, we are
in a situation which is profoundly 
different from the single-state approximation familiar from
the ionization of atoms (initially prepared in their ground state) 
by intense optical fields \cite{Gav92}. 

From (\ref{ionprob}), we can now 
derive a definition of the atomic conductance $g$, in terms of the spectral
information obtained from the diagonalization of $\mathcal H$, as the average
ionization rate at $t\simeq 0$ \cite{BG98}:
\begin{equation}
g \equiv \left. \frac{1}{\Delta} \frac{\rmd}{\rmd t} \Pt \right| _{t \simeq
0} = \frac{1}{\Delta} \sum_{\epsilon}\Gamma_{\epsilon}w_{\epsilon}. 
\label{atomconduct1}
\end{equation}
In order to render $g$ dimensionless, we divided by the average 
level spacing $\Delta$ of the Floquet eigenvalues. 
Furthermore, since in the atomic problem 
there is no incoming particle flux as in the solid-state transmission 
problem, it is reasonable to take the derivative at $t\simeq 0$, in the 
above expression.
Note that the right-hand-side of 
(\ref{atomconduct1}) is strongly reminiscent of Landauer's formula
for the conductance across a disordered sample \cite{Lan70}, 
if we identify the 
Floquet rates $\Gamma_{\epsilon}$ with matrix elements of the transition matrix
in the solid-state problem. Indeed, such an identification can be justified
more formally, as we shall show elsewhere \cite{WB05}.

\section{Numerical results}
\label{atomfluct}

We have now set the scene for our statistical analysis of the atomic 
conductance. To gain a qualitative impression of the phenomenon we are
dealing with, let us first focus on the parameter dependence of the ionization
yield, equation~(\ref{ionprob}), of the initial state $n_0=100$,
for two different values $\ell=0.2$ and $1$, and an interaction time 
$t=300\times 2\pi/\omega$ \cite{KL95}. \Fref{fig1} shows our numerical 
result, within
the interval $\omega_0\in [2.0;2.5]$ (500 equidistant values of $\omega_0$
were found sufficient to resolve all structures of the signal). For $\ell=0.2$,
the ionization yield is typically very small, close to zero, but is locally 
strongly enhanced (by orders of magnitude),
at apparently random values of $\omega_0$. Also for $\ell =1$ the yield
exhibits large fluctuations, however around a clearly finite average value 
larger than zero, and of the same order of magnitude as the average
ionization probability. Both cases are 
reminiscent of conductance fluctuations through disordered solid-state samples,
in the localized ($\ell=0.2$) and in the diffusive (or delocalized, 
$\ell=1$) regime, respectively \cite{WW86,PFWS90}. 
Note that the observed fluctuations
occur on a scale $\delta\omega_0/\omega_0\simeq 10^{-2}$, i.e. $P_{\rm ion}$
is a {\em smooth} function of $\omega_0$ on scales 
$\delta\omega_0/\omega_0\simeq 10^{-3}$ or smaller \cite{BD95b}. For 
$n_0\simeq 60\ldots 100$, this corresponds to a typical frequency window of 
approximately 
$\delta \omega/2\pi\simeq 700\ldots 150\ \rm MHz$, on which the fluctuations
should be detectable, rather than on scales smaller by one order of magnitude,
as considered in \cite{SAKW94}.

To deduce the $\omega_0$-dependence of the atomic conductance from the yield 
displayed in figure~\ref{fig1}, we still need to extract the average level 
spacing $\Delta$ from the raw numerical data. Since not all eigenstates 
$\ket {\epsilon}$ of 
$\mathcal H$ actually contribute to the ionization of $\ket {n_0}$, we 
have to account for the relative weight $w_{\epsilon}$ of the individual 
Floquet eigenstates in our definition of $\Delta$. One way of doing so is to
estimate the number of effectively contributing Floquet states as 
$\exp\left(W_{\rm Shannon}\right)$, via the 
Shannon entropy $W_{\rm Shannon}=-\sum_{\epsilon}w_{\epsilon}\ln w_{\epsilon}$
\cite{BS87}. Another way \cite{BG98} is
to assume that as many Floquet states contribute to the ionization dynamics as 
there are quasi resonantly coupled unperturbed 
states between the atomic initial state and
the continuum threshold, i.e. $N$ (see equation~(\ref{samplelength}) above).
Both estimates consistently provide similar results, i.e., 
\begin{equation}
\Delta\simeq\frac{\omega}{N}\simeq\frac{\omega}{\exp(W_{\rm Shannon})},
\label{estdelt}
\end{equation}
as illustrated in  
figure~\ref{fig2}, in the localized as well as in the delocalized
regime (note that, more precisely, $N>\exp(W_{\rm Shannon})$ in the localized 
regime, and $N<\exp(W_{\rm Shannon})$ in the delocalized regime, what is 
consistent with the interpretation of $N$ as the atomic sample size). 
We checked that this remains true for all values of $n_0$ considered 
hereafter, and that the 
statistical properties of $g$ are insensitive 
to the definition of $\Delta$ we choose, except for an irrelevant offset.
Therefore, all subsequent results are presented with the convention 
$\Delta\equiv\omega/N$.

With this definition, we show the $\omega_0$-dependence of the atomic 
conductance in figure~\ref{fig3}, for two values of $n_0=40,100$, as well as
for the two values  
of the localization parameter already employed in figure~\ref{fig1}. 
Clearly, the erratic fluctuations of $P_{\rm ion}$ carry over to
the atomic conductance. Note that in the localized regime ($\ell =0.2$) the 
fluctuations manifest on a {\em logarithmic} scale ($\ln g$ is plotted vs. 
$\omega_0$ on the left column of figure~\ref{fig3}), whereas the fluctuations 
of $g$ are of the same order of magnitude as its average value, in the 
delocalized regime
($\ell =1$, right column of figure~\ref{fig3}, $g$ is plotted vs. 
$\omega_0$).
Furthermore, the scale $\delta\omega_0/\omega_0$ of the fluctuations clearly 
becomes finer as the principal quantum number $n_0$ is increased from $n_0=40$
to $n_0=100$, an observation which is consistent with the increased density 
of states as one approaches the ionization threshold.

It should be stressed here that such large fluctuations, especially 
those in the 
localized regime, are truely remarkable, since they manifest in a quantity
which represents a weighted {\em average} over the entire Floquet spectrum,
according to equation (\ref{atomconduct1}). 
In an experiment with, say, $n_0=80$, and a carrier frequency of approximately 
$\omega/2\pi\simeq 30\ \rm GHz$, a detuning of approximately $100\ \rm MHz$ can
enhance the ionization yield from virtually zero to more than $10\%$!
It has been shown earlier \cite{BD95b,ZDB98} that specific, {\em individual} 
Floquet eigenstates of the 
atom in the field may exhibit large scale, erratic fluctuations of their 
ionization rates under changes of some control parameter, for instance of
$\omega_0$. 
Here, it is the conspiracy of the distribution of the weights $w_{\epsilon}$
and rates $\Gamma_{\epsilon}$ over the entire spectrum which produces a similar
effect!

Let us now proceed to a first quantitative test of photonic localization
theory. If the atomic localization parameter $\ell$ defined in 
(\ref{lparameter}) indeed plays the same crucial role as in the solid-state 
problem, then $\langle\ln g\rangle$ should decrease 
linearly with increasing $\ell^{-1}$,
by virtue of (\ref{lparameter},\ref{atcond}). \Fref{fig4} shows our 
numerical result, for different values of $n_0$. Apparently, the solid-state 
prediction is almost perfectly followed for the largest principal quantum 
number (i.e., the largest individual values of $\langle\xi\rangle$ and $N$ in 
(\ref{locallength},\ref{samplelength})). On the other hand, the smaller $n_0$,
and, hence, the smaller $\langle\xi\rangle$ and $N$, the larger are the 
deviations from the linear dependence. 
This, however, can be readily understood since the deviations
systematically (for all values of $n_0$) occur for localization parameters
(small $\ell$, large $\ell ^{-1}$) 
which correspond to average localization lengths $\langle\xi\rangle<3$ 
(down
to $\langle\xi\rangle\simeq 0.6\ldots 0.75$, for $\ell=0.1$ and $n_0=40$). 
Then, according to the simple
picture developed in the introduction, no more than two bound states of the 
atom are efficiently coupled by the driving field, and it does not make any
sense to speak of an electronic wave function which is exponentially localized
over quasi resonantly coupled bound states on the energy axis. As a matter of
fact, it is rather surprising that the linear behaviour is observed in 
figure~\ref{fig4} for values of $\langle\xi\rangle$ as small as 3 or 4,
since the assumptions \cite{CC87} for the derivation of (\ref{locallength})
imply $\langle\xi\rangle\gg 1$.

A further quantitative test of the analogy between atomic and 
solid-state transport problem is the statistical distribution of the atomic 
conductance sampled over different values of $\omega_0$, for fixed $\ell$.
Figures \ref{fig5} and \ref{fig7} show histograms of $\ln g$, 
for 
$n_0=40,\ 100$, respectively, and $\ell=0.1\ldots 0.5$ (for each histogram,
approx. $10000\ldots 60000$ resonances with nonvanishing weights
$w_{\epsilon}$ contribute to $500$ values of $g$). Systematically, the 
lognormal fits of the histograms obtained from our data improve as $n_0$ is 
increased, in particular in the wings of the distributions. For small $\ell$, 
low values of 
$n_0$ tend to induce a sharp cut-off at small conductances, what we attribute 
once more to the increasing ``granularity'' of the ionization process as 
$n_0$ decreases. For principal quantum numbers $n_0\geq 70$, however, the 
distributions of the atomic conductance are well fitted \cite{wimda} 
by the lognormal  
prediction derived from the Anderson model.

For given $n_0$ and growing $\ell$, 
the distribution of $\ln g$ shifts to larger values, as 
visible, e.g., in figures~\ref{fig7} and \ref{fig9}, 
for $n_0=100$, and also from
figure~\ref{fig4}. Furthermore, as already could be expected from the
comparison of the $\omega_0$-dependence of the atomic conductance for
$\ell=0.2$ and $\ell=1$ (figure~\ref{fig3}), the lognormal fit ceases to be a
good  
approximation of the histograms for too large values of $\ell$. As evident from
figure~\ref{fig9}, the distribution starts to get asymmetric at $\ell=1$, and 
is clearly not lognormal any more for $\ell=2$. 
This transition from the localized
to the delocalized (or diffusive \cite{CC87,KM93}) regime is even more 
pronounced in 
figure~\ref{fig10}, where we plot the histogram of $g$ rather than of $\ln
g$:  
with $\ell$ increasing from $0.75$ to $2$, the distribution shifts to larger
values, broadens, and develops a large gap at $g=0$. Still, a Gaussian
distribution of $g$ as observed for diffusive transport 
in the solid-state transmission problem 
\cite{PZIS90} cannot be established here. 

Finally, we examined the variances of $\ln g$ as obtained from our numerical 
data, for different values of $n_0$. 
The result is shown in figure~\ref{fig11}. Whereas localization theory
suggests a linear dependence ${\rm Var}(\ln g)\sim -\langle \ln g\rangle$ 
\cite{PZIS90}, our data appear to support this expectation 
only within a finite interval of $\langle \ln g\rangle$, which furthermore
depends on $n_0$, and can be roughly confined by the limits 
$-12\leq \langle \ln g\rangle\leq -7$. For small values of $\langle \ln g
\rangle$, the 
variance systematically drops faster than linearly, and it turns out that 
its overall dependence on the average conductance is best fitted by a 
quadratic law, for all $n_0$. Once again, we attribute
this deviation from  
the solid-state picture to the finite size effect which already manifested 
itself in the dependence of $\langle \ln g \rangle$ on $\ell^{-1}$, and in the 
distribution of $\ln g$, for small values of $\ell$. For too small localization
lengths, the distribution of $\ln g$ is not lognormal any more. Deviations
notably occur in the wings, and we cannot expect a linear variation of 
${\rm Var}(\ln g)$ in this parameter regime. On the other extreme, at large 
localization parameters, the variance saturates, in accordance with our 
observations in figures~\ref{fig3}, \ref{fig4}, and \ref{fig10}, as well as
with general expectations for diffusive transport in disordered solids
\cite{KM93}. 

\section{Conclusions}
\label{conclusions}
To summarize, we can now respond to the questions we formulated at the 
beginning of this paper. Our above results demonstrate that 
the essential statistical features of the conductance across an
Anderson-localized solid-state sample indeed carry over to the fully
deterministic 
ionization process of quasi 
one-dimensional hydrogen Rydberg states under microwave
driving, where dynamical chaos substitutes disorder. In response to our 
first introductory question, we do observe important amendments to the 
solid-state picture (which implicitly always assumes a localization length
much larger than the typical distance between neighbouring lattice 
sites) imported
to the atomic realm by photonic localization theory. These amendments directly
originate -- via the atomic sample size $N$, equation~(\ref{samplelength}) --
in the finite size of $\hbar_{\rm eff}\sim n_0^{-1}$. However, they
do neither affect the monotonous decrease of the mean 
atomic conductance
with the inverse of the localization parameter (figure~\ref{fig4}), 
nor the large-scale fluctuations of $g$, for different realizations of
$\ell$.  

A rough estimate allows an answer to
our second question: Figures \ref{fig9}--\ref{fig11} suggest that the
transition from localized to diffusive transport sets in at $\ell\geq
0.5$. Furthermore, we have seen that deviations from the exponential
dependence of $\langle\ln g\rangle$ on $\ell^{-1}$ get manifest for
$\langle\xi\rangle <3$. Considering the monotonous decrease of $\langle\ln
g\rangle$ with decreasing $\ell$ (figure \ref{fig4}) as the most robust
signature of localization, we therefore require that the minimum value of
$n_0$ permits localization lengths $\langle\xi\rangle > 2$ at $\ell=0.5$. By
virtue of (\ref{locallength}-\ref{lparameter}), this implies 
$n_0\simeq4\omega_0\langle\xi\rangle
> 8\omega_0\simeq 16\ldots 20$ (with $\omega_0 = 2.0 \ldots 2.5$), 
and is consistent with earlier numerical results \cite{BDG95} for
smaller values of $n_0 \simeq 23$. On the other hand, however, the
exponentially enhanced 
fluctuations of the atomic conductance, as the actual, quantitative 
fingerprint of Anderson
localization in our driven atomic system, do fully prevail only for $n_0>70$
(figures~\ref{fig5}-\ref{fig9}, see also \cite{wimda}).

The reply to our third question directly follows from our discussion of 
figure~\ref{fig11} above, which suggests a smooth transition from lognormally
distributed atomic conductances for sufficiently large localization lengths 
and principal quantum numbers, to more coarse-grained distributions with
larger variances. Taking into account 
the perturbative coupling limit defined by small values of $\langle \xi
\rangle \simeq 2$, it should be possible to derive an 
approximate analytical expression for the general behaviour of 
${\rm Var}(\ln g)$ with $\langle \ln g\rangle$.  

\begin{figure}
\centerline{\epsfig{figure=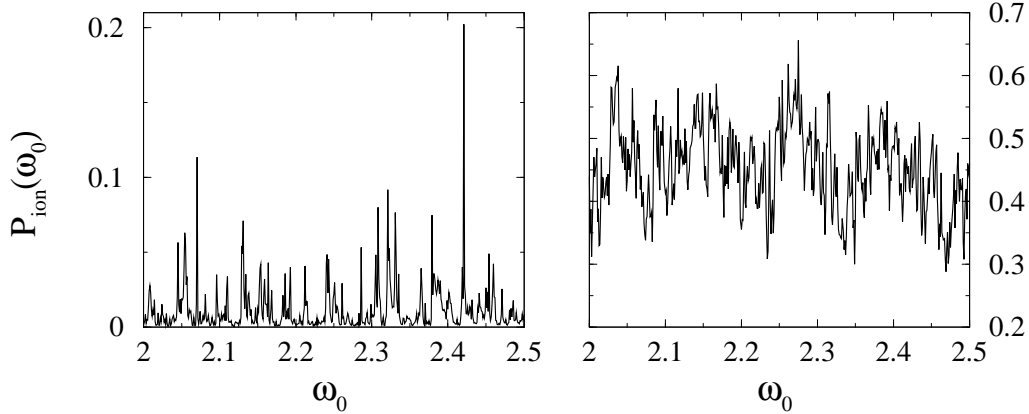,width=6cm,angle=270}}
\caption{Ionization yield $P_{\rm ion}(t)$ vs. the scaled frequency 
$\om$, for an initial principal quantum number $n_0=100$, and 
fixed interaction time $t=300\times 2\pi/\omega$. 
In the localized regime ($\ell =0.2$, left), the
ionization yield is close to zero, 
amended by erratic fluctuations. In the delocalized regime
($\ell =1$, right), the average yield is clearly finite, with erratic
fluctuations of the same order of magnitude.}
\label{fig1}
\end{figure}

\begin{figure}
\centerline{\epsfig{figure=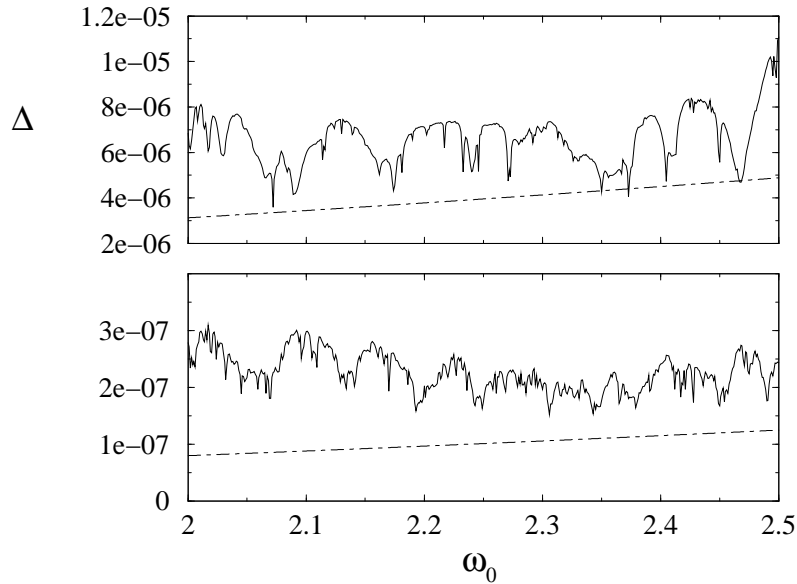,width=8cm,angle=270}}
\caption{Comparison of the two estimates of the average level spacing
$\Delta$ (equation (\protect\ref{estdelt})), 
as a function of the scaled frequency $\om$, for 
constant localization
parameter $\ell=0.2$. The full curves show the
spacings deduced from the Shannon entropy $W_{\rm Shannon}$, which in the
localized regime ($\langle\xi\rangle < N$) are typically %by a factor of two 
larger than those given by the simple estimate $\omega/N$
(dashed-dotted lines). 
As the quantum number $n_0$ of the initial Rydberg state increases from 
40 (top) to 100 (bottom), the absolute values of the spacings decrease. 
The statistical properties of the atomic conductance (\ref{atomconduct1})
discussed hereafter turn out to be independent of the definition of
$\Delta$, except for an irrelevant offset in the 
statistical distributions.}
\label{fig2}
\end{figure}

\begin{figure}
\centerline{\epsfig{figure=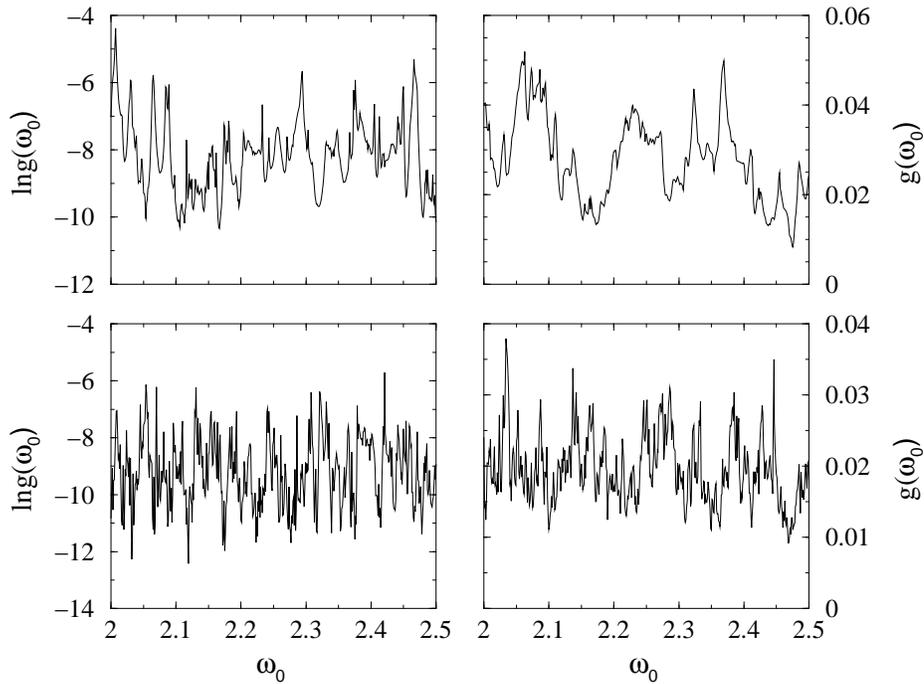,width=9.5cm,angle=270}}
\caption{Atomic conductance vs. scaled frequency $\om$, for localization
parameters $\ell =0.2$ (left), $1$ (right), and
for initial atomic Rydberg states $n_0=40$ (top), $n_0=100$ (bottom),
respectively. The semi-logarithmic plots for the localized case $\ell =0.2$
(left column) clearly exhibit huge fluctuations over several orders of
magnitude, a characteristic feature of quantum transport 
in the presence of Anderson localization (see text). In the delocalized 
regime ($\ell=1$, right column) the amplitude of the fluctuations is 
strongly reduced (note the linear scale) and comparable to the average 
conductance.}  
\label{fig3}
\end{figure}

\begin{figure}
\centerline{\psfrag{1/l}{\bf \large $ 1/\ell$}
\psfig{figure=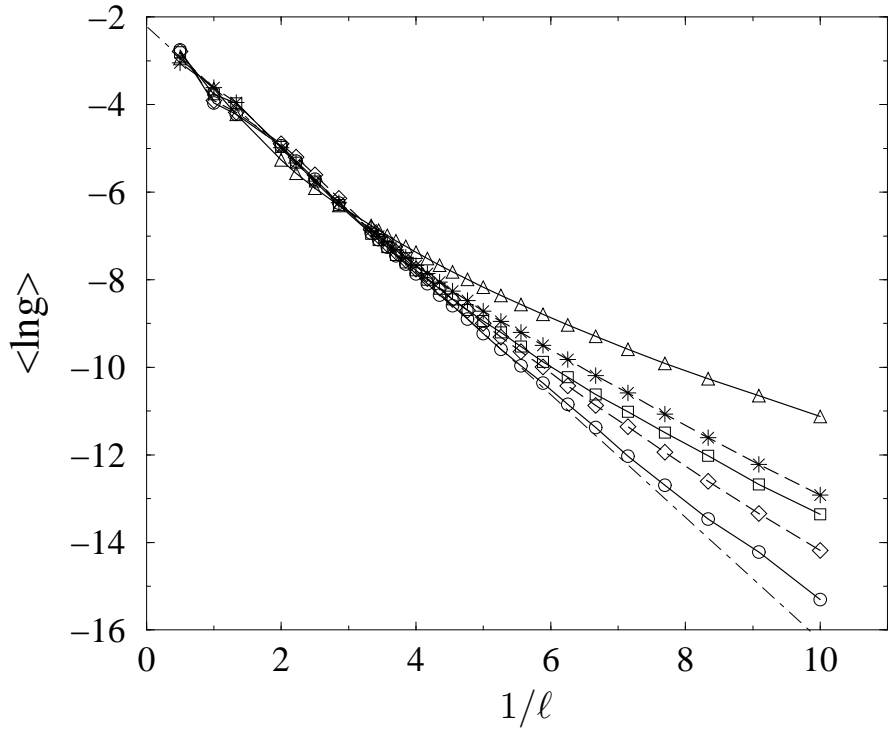,width=10cm,angle=270}}
\caption{Average value of $\ln g$ vs. the inverse localization 
parameter $\ell^{-1}$, for principal quantum numbers $n_0=40\
(\opentriangle)$, $60\ (\star)$,
$70\ (\opensquare)$, $90\ (\opendiamond)$, $100\ (\opencircle)$ of the atomic
initial state, and
$\ell=0.1\ldots 2$. $\langle\ln g\rangle$ was obtained from sampling 
$g$ for 500 equidistant values of $\omega_0\in[2.0;2.5]$, at fixed $\ell$ (see
equation~(\ref{lparameter})). 
For $n_0=100$ we observe an almost perfectly 
linear dependence, in agreement with (\ref{atcond}) and, hence, with the 
Anderson picture. As $n_0$ decreases, $\langle\xi\rangle$ and $N$ decrease at
fixed $\ell$, and a clear deviation from an exponential decrease of 
$\langle\ln g\rangle$ with $\ell^{-1}$ is systematically observed for 
$\langle\xi\rangle <3$.}  
\label{fig4}
\end{figure}

\begin{figure}
\centerline{\psfig{figure=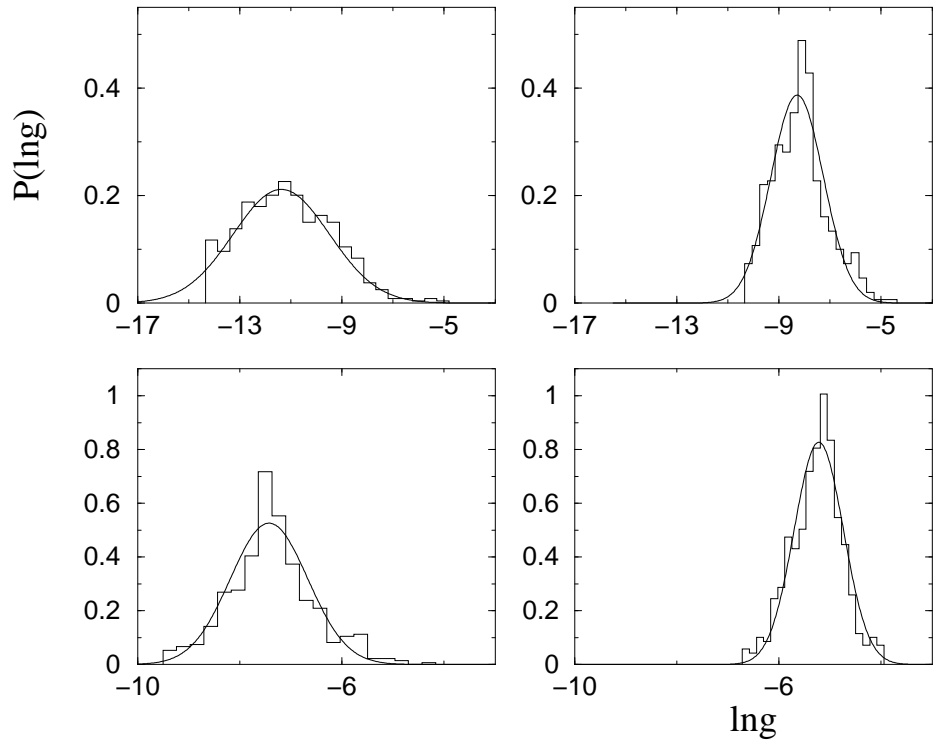,width=10cm,angle=270}}
\caption{Distributions of the logarithm of the atomic conductance $\ln g$
sampled over 500 equidistant values of $\omega_0
\in [2.0;2.5]$, for each value of the localization parameter $\ell =0.1$, 
$0.2$, $0.25$, $0.5$, (top left to bottom right),
and $n_0=40$. The thick lines show
the best fit to a 
normal distribution which is expected on the grounds of Anderson localization
theory. The histograms shift to higher values of $\ln g$, with decreasing 
widths 
as $\ell$ grows. Clearly, for this smallest $n_0$-value employed in our
calculations,
there are considerable deviations from the expected lognormal behaviour.}
\label{fig5}
\end{figure}

\begin{figure}
\centerline{\psfig{figure=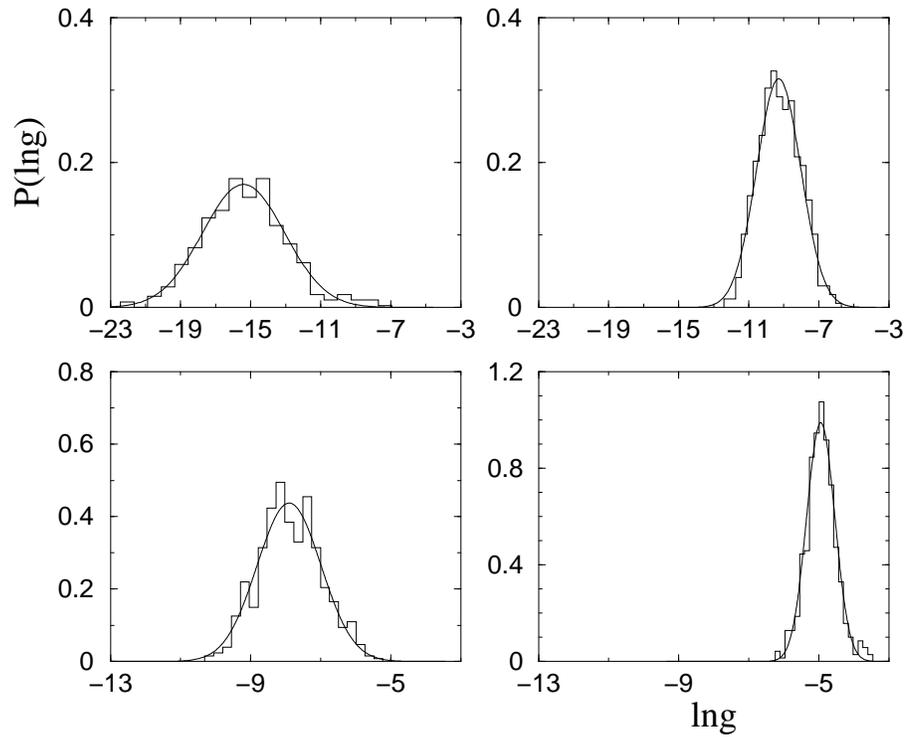,width=10cm,angle=270}}
\caption{Histograms of $\ln g$ fitted by a normal distribution (thick line), 
for $n_0=100$ and the same 
localization parameters  
and sampling interval as in figure~\ref{fig5}.
The agreement with the lognormal prediction implied by Anderson localization
theory is essentially perfect now, at sample sizes and average localization 
lengths $N\simeq15\ldots 19$ and $\langle\xi\rangle\simeq 2\ldots9$, 
respectively, for
$\ell=0.1\ldots 0.5$.}
\label{fig7}
\end{figure}

\begin{figure}
\centerline{\psfig{figure=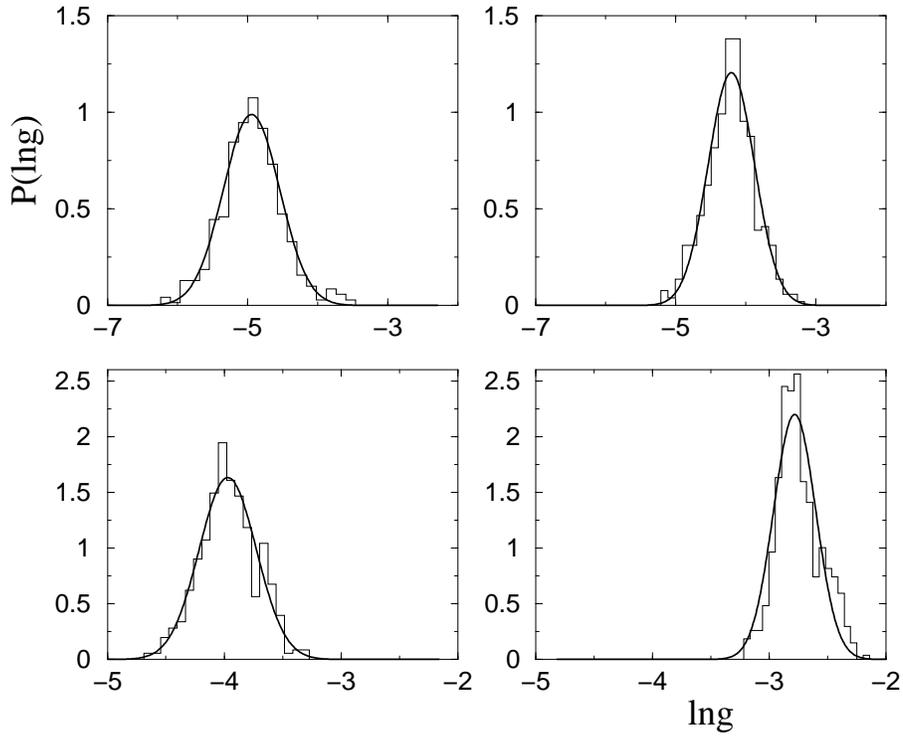,width=10cm,angle=270}}
\caption{Distributions of the logarithm of the atomic conductance for
$n_0=100$ and localization parameters $\ell=0.5\ldots 2$ (top left to bottom
right). The distributions
shift to larger values of $\ln g$ and get narrower as $\ell$ is
increased. At $\ell =2$, a clear deviation 
from the lognormal fit is observed, which reveals the transition to the
delocalized regime. Similar results are obtained over the entire range
$n_0=40\ldots 100$.} 
\label{fig9}
\end{figure}

\begin{figure}
\centerline{\psfig{figure=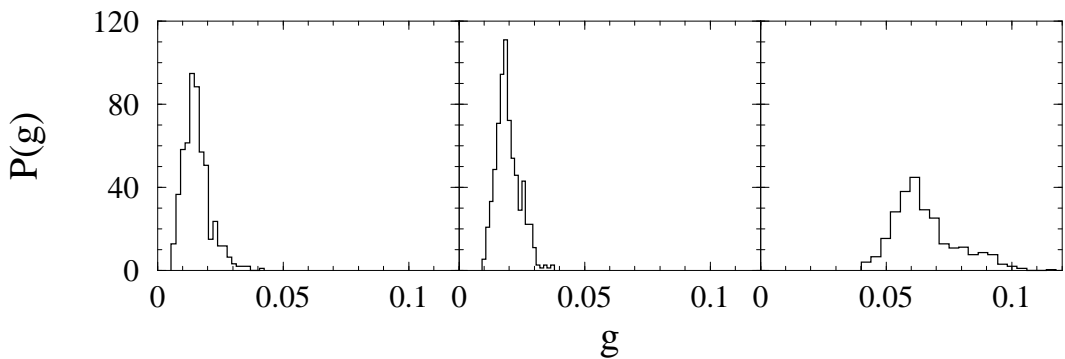,width=5cm,angle=270}}
\caption{Distributions of the atomic conductance of $n_0=100$, for large 
localization parameters $\ell=0.75$ (left), $1$ (middle), and $2$ (right), 
as in figure~\ref{fig9},
but on a linear scale. The broadening of the distribution with increasing
$\ell$, together with the widening gap at $g=0$, indicates the transition
to the delocalized (diffusive) regime.}
\label{fig10}
\end{figure}

\begin{figure}
\centerline{\psfig{figure=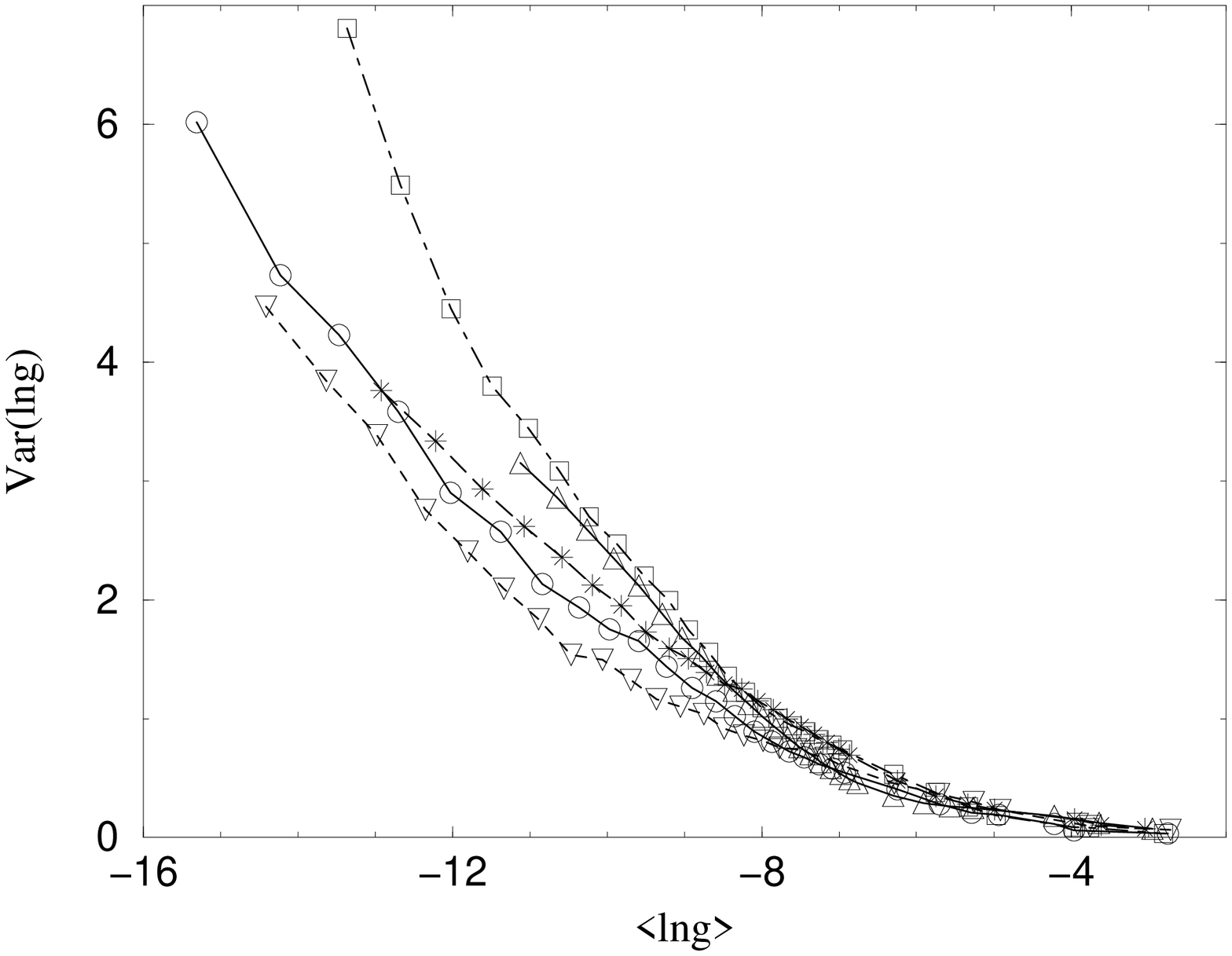,width=10cm,angle=270}}
\caption{Variances of the atomic conductance $\ln g$ vs.~its average
$\langle\ln g\rangle$,
for $n_0=40\ (\opentriangle)$, $60\ (\star)$, $70\
(\opensquare)$, $80\ (\opentriangledown)$, and $100\ (\opencircle)$, 
respectively. 
The localization
parameter $\ell$ increases from left to right ($\ell= 0.1\ldots 2.$). Each
data point was obtained by sampling $g$ over 500 equidistant values of 
$\om \in [2.0;2.5]$, for given $\ell$.}  
\label{fig11} 
\end{figure}

Finally, we verified that the fluctuations displayed in figure~\ref{fig3}
remain essentially unaffected if we average over a finite window of $F_0$
(and, hence, of $\ell$), at any given value of $\omega_0$. We assumed a 
relative error of $\delta F_0/F_0 \leq 5\%$ in the experimental
calibration of the field amplitude experienced by the atoms, which is state of
the art in laboratory experiments \cite{ABMW91,NGG00}. Therefore,  
exponentially large fluctuations 
of the atomic conductance in the localized regime 
should be observable via measurements of the ionization
yield at sufficiently short interaction times (see figures~\ref{fig1} and 
\ref{fig3}). Importantly, such experiments must be performed {\em below}
the ionization (i.e., delocalization) threshold, as immediately apparent 
from figure~\ref{fig1}. Since, so far, most experimental evidence in 
support of dynamical
localization 
in periodically driven atoms is based on measurements of the ionization
threshold \cite{GSMKR88,BCGS89,ABMW91,NGG00}, which doesn't prove more than a
monotonous decrease  
of the average conductance with decreasing $\ell$, the crucial experimental
test of Anderson localization in driven atomic systems is yet to be performed.

\ack
We thank Uzy Smilansky for his warm hospitality during a visit at the 
Weizmann Institute of Science, where part of this work was accomplished. 
Financial support by the Minerva Foundation (AB), and by the
Studienstiftung des dt. Volkes (SW) is gratefully acknowledged. 
We also are grateful to Jakub Zakrzewski, for valuable discussions and for 
making available his Gaussian
fit routine.

\section*{References}

\begin{thebibliography}{32}

\bibitem{houches95}
{\em Mesoscopic quantum physics}, edited by Akkermans E, Montambaux G,
Pichard J~L, and Zinn-Justin J (North-Holland, Amsterdam, 1995).

\bibitem{And58}
Anderson P~W, {\em Phys.~Rev.} {\bf 109},  1492  (1958).

\bibitem{KM93}
Kramers B and MacKinnon A, 
{\em Rep.~Prog.~Phys.} {\bf 56},  1469  (1993).

\bibitem{CCFI79}
Casati G, Chirikov B~V, Ford J, and Izrailev F~M,  in {\em {Stochastic
  Behavior in Classical and Quantum Hamiltonian Systems}}, 
edited by G. Casati and J. Ford (Springer-Verlag, Berlin, 1979), p.\ 334.

\bibitem{FGP82}
Fishman S., Grempel D~R, and Prange R~E,
{\em Phys. Rev. Lett.} {\bf 49},  509  (1982).

\bibitem{CCS84}
Casati G, Chirikov V, and Shepelyansky D L, 
  {\em Phys.~Rev.~Lett.} {\bf 53},  2525  (1984).

\bibitem{CC87}
Casati G, Chirikov V, Shepelyansky D~L, and Guarneri I,
  {\em Phys.~Rep.} {\bf 154},  77  (1987).

\bibitem{SJNS00}
Starykh Q~A, Jacquod P~R, Narimanov E~E, and Stone A~D, 
  {\em Phys.~Rev.~E} {\bf 62},  2078  (2000).

\bibitem{DB94}
Delande D and Buchleitner A, 
  {\em Adv.~At.~Mol.~Phys.} {\bf 34},  85  (1994).

\bibitem{LW97}
Leitner D~M and Wolynes P~G, 
  {\em Chem.~Phys.~Lett.} {\bf 276},  289  (1997).

\bibitem{BG98}
Buchleitner A, Guarneri I, and Zakrzewski J, 
{\em Europhys. Lett.} {\bf 44},  162  (1998).

\bibitem{PT94}
Press W~H, Teulosky S~A, Vetterling W~T, and Flannery B~P, {\em
  {Numerical Recipes in FORTRAN: the art of scientific computing}}, 2 ed.
  (Cambridge University Press, Cambridge, 1994).

\bibitem{CGS90}
Casati G, Guarneri I, and Shepelyansky D~L, {\em Physica A} {\bf 163},  205
  (1990).

\bibitem{GSMKR88}
Galvez E~J, Sauer J~E, Moorman L, Koch P~M, and Richards D,
  \PRL {\bf 61},  2011  (1988).

\bibitem{BCGS89}
Bayfield J~E, Casati G, Guarneri I, and Sokol D~W,
  {\em Phys.~Rev.~Lett.} {\bf 63},  364  (1989).

\bibitem{BBGSSW91}
Bl\"umel R, Buchleitner A, Graham R, Sirko L, Smilansky U, and Walther H,
  {\em Phys.~Rev.~A} {\bf 44},  4521  (1991).

\bibitem{ABMW91}
Arndt M, Buchleitner A, Mantegna R~N, and Walther H,
  \PRL {\bf 67},  2435  (1991).

\bibitem{NGG00}
Noel W~M, Griffith W~M, and Gallagher T~F, {\em Phys.~Rev.~A} {\bf 62},
063401  (2000). 

\bibitem{AALR79}
Abrahams E, Anderson P~W, Licciardello D~C, and Ramakrishnan T~V, 
  \PRL {\bf 42},  673  (1979); 
Efetov K~B, {\em Adv. Phys.} {\bf 32}, 53
(1983); Stone A~D, in \protect\cite{houches95}, and references therein.

\bibitem{PZIS90}
Pichard J~L, Zanon N, Imry Y, and Stone A~D,
  {\em J. Phys. (Paris)} {\bf 51},  587  (1990).

\bibitem{BDG95}
Buchleitner A, Delande D, and Gay J~C, 
  {\em J. Opt. Soc. Am. B} {\bf 12},  505  (1995).

\bibitem{KL95}
Koch P~W and van Leeuwen K, {\em Phys.~Rep.} {\bf 255},  289  (1995).

\bibitem{BD97}
Buchleitner A and Delande D, {\em Phys. Rev. A} {\bf 55},  R1585  (1997).

\bibitem{Gav92}
Gavrila M,  in {\em Atoms in Intense Laser Fields}, Vol.~1 of {\em Advances in
  atomic, molecular, and optical physics, {\rm supplements}}, edited by M.
  Gavrila (Academic Press, Boston, 1992), p.\ 435.

\bibitem{Lan70}
Landauer R, {\em Philos.~Mag.} {\bf 21},  863  (1970).

\bibitem{WB05}
Wimberger S and Buchleitner A (unpublished).

\bibitem{WW86}
Washburn S and Webb R~A, {\em Adv.~Phys.} {\bf 35},  375  (1986).

\bibitem{PFWS90}
Popovic D, Fowler A~B, Washburn S, and Stiles P~J, 
  {\em Phys.~Rev.~B} {\bf 42},  1759  (1990).

\bibitem{BD95b}
Buchleitner A and Delande D, {\em Chaos, Solitons \& Fractals} {\bf 5},
  1125  (1995).

\bibitem{SAKW94}
Sirko L, Arndt M, Koch P~M, and Walther H,
  {\em Phys.~Rev.~A} {\bf 49},  3831  (1994).

\bibitem{BS87}
Bl\"umel R and Smilansky U, {\em Phys.~Rev.~Lett.} {\bf 58},  2531  (1987).

\bibitem{ZDB98}
Zakrzewski J, Delande D, and Buchleitner A, 
  {\em Phys.~Rev.~E} {\bf 57},  1458  (1998).

%ab adds
\bibitem{wimda}
Wimberger S, Diploma thesis, 
Ludwig-Maximilians-Universit\"at M\"unchen, M\"unchen, 2000.
%ab

\end{thebibliography}
\end{document}